\documentclass[twocolumn,showpacs,eqsecnum,floatfix,superscriptaddress]{revtex4}

\usepackage{graphicx}
\usepackage{dcolumn}
\usepackage{bm}
\usepackage{amsmath}

\usepackage{xcolor}
\usepackage{soul}
\usepackage{graphicx}
\usepackage{dcolumn}
\usepackage{bm}
\def\lesssim{\ \raise.3ex\hbox{$<$}\kern-0.8em\lower.7ex\hbox{$\sim$}\ }
\def\gesim{\ \raise.3ex\hbox{$>$}\kern-0.8em\lower.7ex\hbox{$\sim$}\ }

\begin{document}
\title{Projectile Trajectory of Penguin's {Faeces} and Rectal Pressure Revisited}
\author{Hiroyuki Tajima}
\affiliation{Department of Mathematics and Physics, Kochi University, Kochi 780-8520, Japan}
\author{Fumiya Fujisawa}
\affiliation{Katsurahama Aquarium, Kochi 781-0262, Japan}
\date{\today}
\begin{abstract}
We discuss a trajectory of penguins' faeces after the powerful shooting due to their strong rectal pressure. 
Practically, it is important to see how far faeceses reach when penguins expel them from higher places.
Such information is useful for keepers to avoid the direct hitting of faeceses. 
We estimate the upper bound for the maximum flight distance by solving the Newton's equation of motion.
Our results indicate that 
the safety zone should be 1.34 meters away from a penguin trying to poop in typical environments. 
In the presence of the viscous resistance, the grounding time and the flying distance of faeces can be expressed in terms of Lambert {\it W} function.
Furthermore, we address the penguin's rectal pressure within the hydrodynamical approximation combining Bernoulli's theorem and Hagen-Poiseuille equation for viscosity corrections.
We found that the calculated rectal pressure is larger than the estimation in the previous work.  
\end{abstract}
\maketitle
\section{Introduction}
Penguins, which are aquatic birds living mostly in the Southern Hemisphere~\cite{Borboroglu},
strongly shoot their faeceses towards their rear side~\cite{MeyerRochow}. 
It is believed that this is because penguins avoid getting the faeces on themselves as well as the nest.
Although such a tendency is not limited to penguins and can be found in the case of other birds,
these bombings sometimes embarrass keepers under breeding environments like an aquarium.
It is practically important to know how far their faeceses reach from the origin.
Such information would save keepers from the crisis.
It would also be helpful for a newcomer guidance for keepers to avoid such an incident.
\par
The flying distance of penguin's faeces reaches about 0.4~m even on the ground.
Since a typical height of a Humboldt penguin is given by 0.4~m,
this distance corresponds to the situation that if a human being whose height is 1.7~m tries to evacuate his/her bowels, the object could fly to 1.7~m away.
Therefore, one can immediately understand that penguin's rectal pressure is relatively much strong compared to that of a human kind.  
In the pioneering work of Ref.~\cite{MeyerRochow}, it is reported that this actual pressure could range from 10 kPa to 60 kPa for relevant values of the faeces viscosity and the radius of the bottom hole. 
\par
\begin{figure}[t]
\begin{center}
\includegraphics[width=7cm]{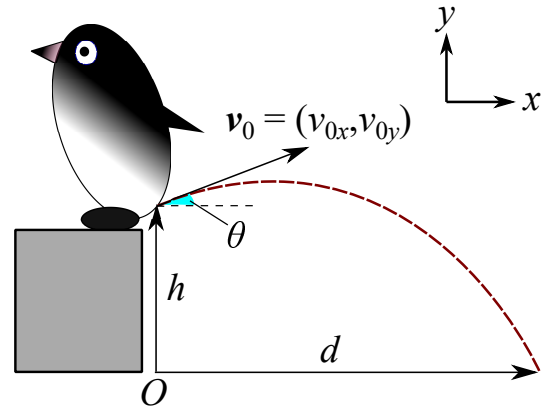}
\end{center}
\caption{Configuration for a penguin trying to defecate towards his/her rear side.
A penguin stands on the rock with the height $h$ from the ground.
We parameterize the ejection angle $\theta$ with the initial velocity $\bm{v}=(v_{0x},v_{0y})$ We estimate a flying distance $d$ of the faeces from the origin $O$.
}
\label{fig1}
\end{figure}

However, in Ref.~\cite{MeyerRochow}, the projectile trajectory of the faeces is not taken into account but the horizontal distance is employed to estimate the fluid volume.
In general, the actual trajectory would be longer than the horizontal distance during the projectile motion as shown in Fig.~\ref{fig1}.
In addition, since the ejection angle is not always horizontal and penguins sometimes shoot them out from higher place under the breeding environment,
it is important to consider such a motion of faeces in a more general manner.
The rectal pressure is regarded as an impulse force to accelerate the faeces up to the initial velocity in Ref.~\cite{MeyerRochow}. 
For non-viscous fluids, we can use Bernoulli's theorem which is related to the energy conservation. 
To estimate the viscosity effects on the rectal pressure, Ref.~\cite{MeyerRochow} assumed the Hagen-Poiseille flow in the air, which is laminar flow of Newtonian liquid in a cylindrical pipe geometry.  
A similar approach for uniaxial urinary flow was employed to estimate the duration of urination of various animals in Refs.~\cite{Yang,Yang2}.
\par 
In this work, we calculate the maximum flying distance of penguin's faeces from a high place, which is relevant for several breeding environments.
Such a projectile trajectory is described by Newton's equation of motion.
We assume that the upper bound for the flying distance can be obtained by the equation of motion in the absence of the air resistance.
Moreover, we revisit the rectal pressure by using Bernoulli's theorem and the Hagen-Poiseille equation~\cite{Marine} to estimate the mechanical contributions of non-viscous flow and the viscosity correction during the flow in the stomach and in the air, respectively. 
\par
This paper is organized as follows.
In Sec.~\ref{setup}, we explain our setup for the projectile motion of penguin's faeces and the evaculation of them from their intestines.
The latter is used for the estimation of penguin's rectal pressure.
In Sec.~\ref{secNE}, we show Newton's equation of motion for the faeces after the shoot.
We show the maximum flying distance at arbitrary angle and height.
In Sec.~\ref{secNEVR}, we discuss effects of the viscous air resistance during the projectile motion.
In Sec.~\ref{secBT}, we calculate the rectal pressure under the assumption that the faeces liquid can approximately regarded as an ideal fluid.
In Sec.~\ref{secHP}, we estimate the additional contribution to the rectal pressure due to the viscosity correction within the Hagen-Poiseuille equation.
Finally, we summarize this paper in Sec.~\ref{secS}.
\section{Setup}
\label{setup}
Before moving to the calculation, 
we explain the configurations and the parameters we consider in this work.
Figure~\ref{fig1} shows the situation where a Humboldt penguin shoots the faeces out from the higher place with the height $h$.  
Here, we employ an initial velocity $|\bm{v}_0|\equiv v_0=2.0$~m/s reported in Ref.~\cite{MeyerRochow} as a typical value.
The ejection angle is denoted by $\theta$, and the gravitational constant is fixed at $g=9.8$~m/s$^2$. 
\par
\begin{figure}[t]
\begin{center}
\includegraphics[width=6cm]{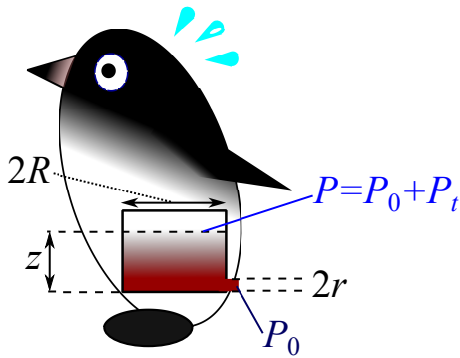}
\end{center}
\caption{Model for penguin's stomach.
We assume that the faeceses are prepared in a fictitious cylindrical tank with the radius $R$.
The depth $z$ is estimated from the fluid volume $V$ after the firing.
The pressure in the stomach is given by $P=P_0+P_t$ where $P_0=1013$ hPa and $P_t$ are the atmospheric pressure and the rectal pressure, respectively.
}
\label{fig4}
\end{figure}
Figure~\ref{fig4} shows the effective model of penguin's stomach.
We employ properties of the faeces used in Ref.~\cite{MeyerRochow},
where the mass density $\rho=1141$~kg/m$^3$, the viscosity $\eta=0.02\sim 0.08$ Pa$\cdot$s are employed.
In our model the radius of bottom hole is given by $r=0.004$~m.
Although it is known that penguin's rectum has a form of a straight tube~\cite{McLelland},
in this work we approximate it as a cylindrical tank with the radius $R$ for simplicity.
The pressure in the stomach is given by $P=P_0+P_t$ where $P_0=1013$~hPa and $P_t$ are the atmospheric pressure in the air and penguin's rectal pressure, respectively.
\section{Newton's equation of motion without air resistances}
\label{secNE}
We consider Newton's equations of motion
\begin{align}
m\frac{d^2x}{dt^2}=0,
\end{align}
\begin{align}
m\frac{d^2y}{dt^2}=-mg,
\end{align}
where $m$ is the total mass of penguin's faeces.
By solving them, one can obtain textbook results
\begin{align}
v_{x}=\frac{dx}{dt}=v_{0x}\equiv v_{0}\cos\theta,
\end{align}
\begin{align}
v_y=\frac{dy}{dt}=v_{0y}-gt\equiv v_{0}\sin\theta -gt,
\end{align}
\begin{align}
x(t)=v_{0}t\cos\theta,
\end{align}
and
\begin{align}
y(t)=h+v_{0}t\sin\theta-\frac{1}{2}gt^2.
\end{align}
We obtain the grounding time $t_{g}$ from $y(t_g)=0$, which reads
\begin{align}
\frac{1}{2}gt_g^2-v_{0y}t_g-h=0.
\end{align}
Thus, one can obtain
\begin{align}
t_{g}&=\frac{v_{0}\sin\theta}{g}+\sqrt{\frac{v_{0}^2\sin^2\theta}{g^2}+\frac{2h}{g}}.
\end{align}
Since the flying distance is given by $d=v_{0}t_g\cos\theta$, we obtain
\begin{align}
d=v_0\cos\theta\left(\frac{v_0\sin\theta}{g}+\sqrt{\frac{v_{0}^2\sin^2\theta}{g^2}+\frac{2h}{g}}\right). 
\end{align}
\begin{figure}[t]
\begin{center}
\includegraphics[width=7cm]{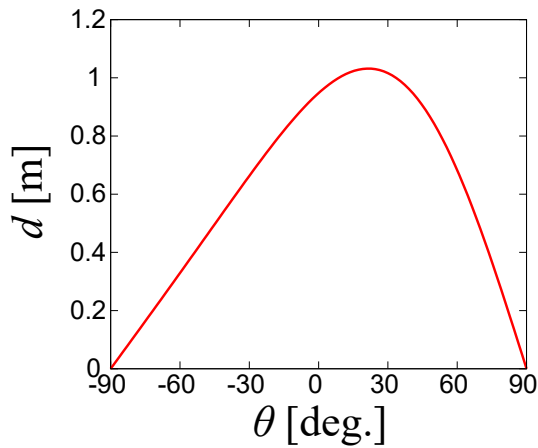}
\end{center}
\caption{ The flying distance $d$ as a function of $\theta$ at $h=1.1$~m.
}
\label{fig2}
\end{figure}
In Fig.~\ref{fig2}, we plot $d$ as a function of $\theta$ at a typical height $h=1.1$~m.
The angle $\theta_{\rm max}$ where $d$ becomes maximum is given by the condition
\begin{align}
\label{eq:dddth}
\left.\frac{\partial d}{\partial \theta}\right|_{\theta=\theta_{\rm max}}&=-v_0\sin\theta\left(\frac{v_0\sin\theta}{g}+\sqrt{\frac{v_{0}^2\sin^2\theta}{g^2}+\frac{2h}{g}}\right)\cr
&+v_0\cos\theta\left(\frac{v_0\cos\theta}{g}+\frac{v_0^2\sin\theta\cos\theta}{\sqrt{\frac{v_{0}^2\sin^2\theta}{g^2}+\frac{2h}{g}}}\right)\cr
&=0.
\end{align}
From Eq.~(\ref{eq:dddth}) we obtain
\begin{align}
\theta_{\rm max}=\sin^{-1}\left[\frac{1}{\sqrt{2\left(1+\frac{hg}{v_0^2}\right)}}\right].
\end{align}
In this regard, in terms of $h$ and $v_0$, the maximum flying distance $d_{\rm max}$ is given by
\begin{align}
d_{\rm max}&=v_0\cos\left[\sin^{-1}\left\{\frac{1}{\sqrt{2\left(1+\frac{hg}{v_0^2}\right)}}\right\}\right]\cr
&\times\left[\frac{v_0}{g}\frac{1}{\sqrt{2\left(1+\frac{hg}{v_0^2}\right)}}+\sqrt{\frac{v_0^2}{2g^2\left(1+\frac{hg}{v_0^2}\right)}+\frac{2h}{g}}\right].\cr
\end{align}
Figure~\ref{fig3} shows $h$-dependence of $d_{\rm max}$ at $\theta=\theta_{\rm max}$.
The inset shows $\theta_{\rm max}$ as a function of $h$.
In the case of $h=1.1$~m which is typical height of rocks in the penguin's area of the  Katsurahama aquarium, we obtain $d_{\rm max}=1.03$~m at $\theta_{\rm max}=21.6^\circ$.
It is slightly longer the case of the horizontal ejection $d(\theta=0)=0.95$~m.
We note that the maximal height of rocks is $h=2.0$~m in the Katsurahama aquarium.
In this case, we obtain $d_{\rm max}=1.34$~m at $\theta_{\rm max}=16.9^{\circ}$ and $d(\theta=0)=1.24$~m.
Since the air resistance in principle shortens $d$ as we discuss in Sec.~\ref{secNEVR}, this value can be regarded as an upper bound for $d_{\rm max}$.
Therefore, we found that penguin keepers should keep the distance being longer than $1.34$~m from penguins trying to eject faeces in the Katsurahama aquarium.  
\begin{figure}[t]
\begin{center}
\includegraphics[width=7cm]{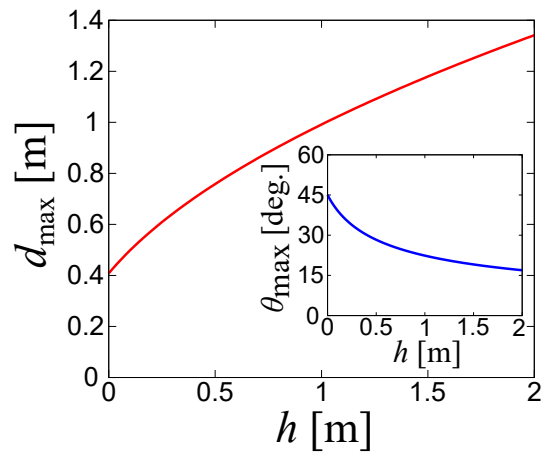}
\end{center}
\caption{The maximum flying distance $d_{\rm max}$ as a function of $h$.
The inset shows the corresponding angle $\theta_{\rm max}$.  
}
\label{fig3}
\end{figure}
\section{Newton's equation of motion with viscous resistance}
\label{secNEVR}
Here, we discuss effects of the air resistance.
For simplicity, we consider the case with viscous resistance which is proportional to $\bm{v}$.
The equations of motion are given by
\begin{align}
m\frac{d^2x}{dt^2}=-kv_x,
\end{align}
\begin{align}
m\frac{d^2y}{dt^2}=-mg-kv_y.
\end{align}
Resulting velocities and distances are given by
\begin{align}
v_x(t)=v_{0x}e^{-\frac{k}{m}t},
\end{align}
\begin{align}
v_{y}(t)=v_{0y}e^{-\frac{k}{m}t}+\frac{mg}{k}(e^{-\frac{k}{m}t}-1),
\end{align}
\begin{align}
x(t)=\frac{mv_{0y}}{k}(1-e^{-\frac{k}{m}t}),
\end{align}
and
\begin{align}
y(t)=h+\frac{m}{k}\left(v_{0y}+\frac{mg}{k}\right)(1-e^{-\frac{k}{m}t})-\frac{mg}{k}t.
\end{align}
We note that by expanding them with respect to $k$, one can reproduce the results in the absence of the resistance.
Since $v_x(t)$ exponentially decreases due to the nonzero $k$,
one can confirm that $d_m{\rm max}$ with $k=0$ gives an upper bound for the distance. 
We note that in this case the grounding time is given by
\begin{align}
t_{g}&=\frac{m}{k}W_0\left(-\left(1+\frac{kv_{0y}}{mg}\right)e^{-\left(1+\frac{kv_{0y}}{mg}+\frac{k^2h}{m^2g}\right)}\right)\cr
&+\frac{m}{k}+\frac{hk}{mg}+\frac{v_{0y}}{g},
\end{align}
where $W_0(x)$ is the main branch of the Lambert {\it W} function.
The distance can be obtained by $d=x(t_g)$.
\par
We also note that the resistance given by $k|\bm{v}|v$ is more realistic in the air~\cite{Parker}.
In such a case, an approximated solution for a low-angle trajectory has also been obtained by using the Lambert {\it W} function~\cite{Belgacem2,Belgacem}.
We note that an analytic solution in this case can be obtained by the homotopy analysis method~\cite{Yabushita}. 
In addition, the viscosity of faeces itself may shorten the flying distance due to the energy dissipation originating from the internal friction.
More sophisticated treatments are required to take such an effect into account.
However, our estimation for the upper bound on $d_{\rm max}$ is robust against these resistance effects.
\section{Bernoulli's theorem and abdominal pressure of penguins}
\label{secBT}
Bernoulli's theorem for non-viscous faeces liquids is given by
\begin{align}
\frac{1}{2}\rho v^2+\rho g z + P= const.
\end{align}
where $z$ is the initial height of liquids in a penguin as shown in Fig.~\ref{fig4}.
While the faeceses are immobile under the abdominal pressure $P_a$ initially, 
they are released with $v_0$ in the air.
Such an assumption gives
\begin{align}
\rho g z + P_a+P_0=\frac{1}{2}\rho v_0^2+P_0,
\end{align}
where $P_0$ is the atmospheric pressure.
At this stage, we do not consider the viscosity correction.
Thus we obtain
\begin{align}
P_a=\frac{1}{2}\rho v_0^2-\rho g z.
\end{align}
The unknown parameter $z$ can be estimated from 
\begin{align}
z=\frac{V}{\pi R^2},
\end{align}
where $V$ is the fluid volume after the ejection.
$R$ is the radius of stomach.
For simplicity, we use $R=0.1$ m.
Here, one may notice that the system is quite similar to the so-called tank orifice with a vena contracta~\cite{Lienhard,Hicks}. 
In such a case, the cross-section area of non-viscous flow after the ejection
is approximately given by $C\pi r^2$, where $C\simeq0.611$ is a typical value of the coefficient of contraction~\cite{Lienhard}.
In this regard, we obtain
\begin{align}
\label{eqV}
V&=C\pi r^2 \ell \cr
&=C\pi r^2\int_0^{t_g} dt\sqrt{v_{0x}^2+(v_{0y}-gt)^2}\cr
&=\frac{C\pi r^2}{2g}\Biggl[v_{0y}v_0-(v_{0y}-gt_g)\sqrt{(gt_g-v_{0y})^2+v_{0x}^2}\Biggr.\cr
&-\left.v_{0x}^2\ln\left(\frac{\sqrt{(gt_g-v_{0y})^2+v_{0x}^2}-gt_g+v_{0y}}{v_0+v_{0y}}\right)\right],\cr
\end{align}
where $\ell$ is the total path of the faeces.
In particular, in the case of $\theta=0$ (namely, $v_{0y}=0$ and $v_{0x}=v_0$),
we obtain
\begin{align}
V&=\frac{C\pi r^2}{2g}\Biggl[gt_g\sqrt{g^2t_g^2+v_{0}^2}\Biggr.\cr
&\left.-v_{0}^2\ln\left(\frac{\sqrt{g^2t_g^2+v_{0}^2}-gt_g}{v_0}\right)\right].
\end{align}
By substituting Eq.~(\ref{eqV}) to
\begin{align}
P_a=\frac{1}{2}\rho v_0^2-\frac{\rho g}{\pi R^2}V,
\end{align} 
we can obtain $P_a\simeq2.3$ kPa at $\theta=0$ and $h=0.2$ m following the previous work~\cite{MeyerRochow}.
It is smaller than the estimated pressure $4.6$ kPa in Ref.~\cite{MeyerRochow} where the initial pressure for the acceleration was obtained by $P_a=\rho v_0^2$.
However, $P_a$ are small compared to the viscosity effect which is addressed in Sec.~\ref{secHP}.
\section{Hagen-Poiseuille equation and viscosity effect}
\label{secHP}
The Hagen-Poiseuille equation gives a relation between an additional pressure and fluid flow rate $Q$ for laminar flow though a cylindrical pipe~\cite{Truedell} as
\begin{align}
Q=\frac{\pi r^4 \alpha}{8\eta},
\end{align}
where $\alpha=\nabla P_b$ and $Q$ can approximately be given by
$\alpha\simeq \frac{P_b}{\ell}$
and
$Q\simeq\frac{V}{t_g}$, respectively.
Therefore, we obtain
\begin{align}
P_b=\frac{8V\ell\eta}{t_g\pi r^4},
\end{align}
which is consistent with Ref.~\cite{MeyerRochow}.
Furthermore, we estimate the additional pressure contributions for the rectal flow as
\begin{align}
P_c=\frac{8Vz\eta}{t_d\pi R^4}\simeq\frac{4Vv_0\eta}{\pi R^4},
\end{align}
where $t_d$ is the flowing time inside the intestine.
We have estimated it as $t_d\simeq\frac{2z}{v_0}\simeq 0.45$ ms.
Even the maximum duration $t_g+t_d\simeq0.20$ s is quite short compared to the urination time $8.2 (M/{\rm kg})^{0.13}$ s obtained in Ref.~\cite{Yang} (where $M$ is the mass of an animal).
We note that a typical value of $M$ for a Hunboldt penguin is about $4$ kg.  
In this way, the total rectal pressure is given by
$P_{t}=P_a+P_b+P_c$.
\par
\begin{figure}[t]
\begin{center}
\includegraphics[width=7cm]{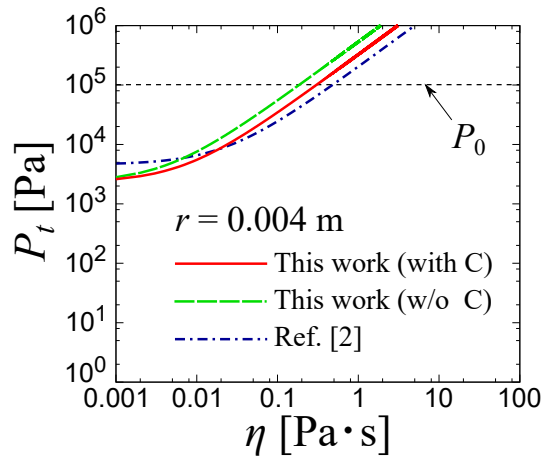}
\end{center}
\caption{Calculated penguin's rectal pressures $P_t$ with and without the coefficient of the contraction $C$, as functions of the viscosity $\eta$.
The dash-dotted curve represents the result of Ref.~\cite{MeyerRochow}.
For comparison, we also show $P_0$. 
}
\label{fig5}
\end{figure}
Figure  \ref{fig5} shows the viscosity dependence of $P_t$ with $r=0.004$ m.
For comparison, we plot the previous result in Ref.~\cite{MeyerRochow}
and our result of $P_t$ without $C$.
While the previous work gives $P_t= 8.56\sim20.6$ kPa for $\eta=0.02\sim 0.08$ Pa$\cdot$s (noting that $\eta$ may change due to the physical condition of penguins),
our results show $P_t=8.75\sim 28.2$ kPa, which is up to $1.4$ times of the previous work.
This is because we consider a longer curved path $\ell$ of flying faeces to estimate $V$ whereas only the horizontal distance was taken account in Ref.~\cite{MeyerRochow}.
We note that if we neglect the contraction as $C=1$, our result gives larger values $P_t=12.9\sim44.7$ kPa since $V$ becomes larger for given $r$.
These results are also larger than the bladder pressure $P_{\rm bladder}=5.2 (M/{\rm kg})^{-0.01}$ kPa in Ref~\cite{Yang}.
This difference may originate from the physical configurations.
Whereas we consider the projectile ejection of faeces, in Ref.~\cite{Yang} a vertical motion of urine was examined.
In addition, since penguins possess the cloacas which combine the digestive and urogenital tracts~\cite{Oliveira},
the visocity and density of penguin's faeces (actually mixing with urine) should be inhomogeneous in the current case. The effect of such an inhomogeneity is left for future work. 
We note that while we use rough approximations for $R$ to calculate $P_a$ and $P_{c}$,
these contributions are found to be small compared to $P_{b}$.
\par
In the end of this section, to realize how strong the estimated penguin's rectal pressure is,
let us demonstrate how far a liquid-like object blasted off from a human being with a severe stomachache flies if his/her rectal pressure is as strong as penguin's one.
Here, we assume that this liquid behaves as a nearly perfect fluid~\cite{KSS} and its density is same as water $\rho=10^3$~kg/m$^3$, for simplicity. 
In this case, we obtain $v_0=7.51$~m/s from $v_0\simeq\sqrt{2P_t/\rho}$ with $P_t=28.2$~kPa.
If the liquid is horizontally launched from the hip with $h=0.85$~m (which is a typical length of legs),
the flying distance can be estimated as $d=3.13$~m, which greatly excesses the typical height of a human being (1.7~m) we mentioned in the introduction.
He/she should not use usual rest rooms.
Although we assume an ideal situation to obtain the numerical value, 
one can easily understand the incredible power of penguin's rectum in this way. 
\section{Summary}
\label{secS}
To summarize, we have discussed a projectile trajectory of penguin's faeces induced by the strong rectal pressure.
We have estimated the upper bound for the maximum flying distance $d_{\rm max}=1.34$~m for the case with a typical initial velocity $v_0=2.0$~m/s of faeces, from the higher place with the height $h=2.0$~m.
In the presence of the air resistance which is proportional to the velocity,
we show that the grounding time and the flying distance of penguin's faeces can be expressed in terms of the Lambert {\it W} function.
\par
Furthermore, based on the hydrodynamic approach, we have revisited penguin's rectal pressure with an approximation combining the Bernoulli's theorem and the Hagen-Poiseulle equation.
The obtained rectal pressure is larger than the previous estimation.
The main difference originates from the treatment of faeces fluid volume flying in the air.
Our results would contribute to further understanding of ecological properties of penguins. 
\par
In this paper, we have used simplified equations of motion to track the trajectory of faeces. 
To obtain a more quantitative results, it is necessary to solve hydrodynamic equations of faeces in the air and in the stomach, which are left for future work.
More sophisticated treatment for the vena contracta would be important.
It is also interesting how complex hydrodynamic behaviors such as turbulence appear during the ejection of faeces. 
\par
We expect that our results would also be a useful example for teaching the classical mechanics in the undergraduate course.
One can recognize that fundamental physics and mathematics we learn in schools describe interesting aspects of various things surrounding our daily life, as demonstrated by T. Terada~\cite{Matsushita}.
\par
\acknowledgments
The authors thank K. Iida, T. Hatsuda, A. Nakamura, and Y. Kondo for useful discussion and the Katsurahama aquarium for the supports and giving us an opportunity to observe Humboldt penguins.

\par


\begin{thebibliography}{99}
\bibitem{Borboroglu} P. B. Borboroglu, and P. D. Boersma, {\it Penguins: Natural History and Conservation.}
(University of Washington Press, 2013).
\bibitem{MeyerRochow} V. B. Meyer-Rochow and J. Gal, Polar Biol. \textbf{27}, 56 (2003). 
\bibitem{Yang} P. J. Yang, J. Pham, J. Choo, and D. L. Hu,
Proc. Natl. Acad. Sci. USA \textbf{111}, 11932 (2014).
\bibitem{Yang2} P. J. Yang, M. LaMarca, C. Kaminski, D. I. Chu, and D. L. Hu,
Soft Matter \textbf{13}, 4960 (2017).
\bibitem{Marine} J. N. Newman, {\it Marine Hydrodynamics}, Cambridge, USA, MIT Press (1977).
\bibitem{McLelland} A. S. King and J. McLelland, {\it Birds: their structure and function.} Bailliere Tindall, London, pp 310–311 (1984).
\bibitem{Parker} G.W. Parker, Am. J. Phys. \textbf{45}, 606, (1977). 
\bibitem{Belgacem2} C. H. Belgacem, Eur. J. Phys. \textbf{35}, 055025 (2014).
\bibitem{Belgacem} C. H. Belgacem, Journal of Taibah University for Science, \textbf{11}, 328 (2017).
\bibitem{Yabushita} K. Yabushita, M. Yamashita, and K. Tsuboi, J. Phys. A: Math. Theor. \textbf{40}, 8403 (2007).
\bibitem{Truedell} C. Truesdell and K. R. Rajagopal, {\it An Introduction to the Mechanics of Fluids.} Birkhuser,
Boston (2009).
\bibitem{Lienhard} J. H. Lienhard and J. H. Lienhard J. Fluids Eng. \textbf{10}, 422 (1984).
\bibitem{Hicks} A. Hicks and W. Slaton, Phys. Teach. \textbf{52}, 43 (2014).
\bibitem{Oliveira} C. A. Oliveira, R. M. Silva, M. M. Santos, and G. A. B. Mahecha,
J. Morphol. \textbf{260}, 234 (2004).
\bibitem{KSS} P. K. Kovtun, D. T. Son, and A. O. Starinets, Phys. Rev. Lett. \textbf{94}, 111601 (2005).
\bibitem{Matsushita} M. Matsushita, Evol. Inst. Econ. Rev. \textbf{6}, 337 (2010).

\end{thebibliography}
\end{document}